\documentclass[manuscript]{acmart}
\AtBeginDocument{%
  \providecommand\BibTeX{{%
    \normalfont B\kern-0.5em{\scshape i\kern-0.25em b}\kern-0.8em\TeX}}}

\usepackage{amsmath}
\usepackage{accents}
\usepackage[tight,nooneline]{subfigure}
\usepackage[ruled,linesnumbered,vlined]{algorithm2e}
\newcommand{\Comment}[1]{\hfill$\triangleright$\ {#1}}

\newcommand{\Eref}[1]{Equation~\eqref{#1}}
\newcommand{\Sref}[1]{Section~\ref{#1}}
\newcommand{\Aref}[1]{Algorithm~\ref{#1}}
\newcommand{\fref}[1]{Fig.~\ref{#1}}

\newcommand{\topn}{\mbox{top-$n$\ }}
\DeclareMathOperator{\simi}{sim}

\newcommand{\QR}[1]{\text{QR}\left({#1}\right)}
\newcommand{\bvec}[1]{\mathbf{#1}}
\newcommand{\R}{\mathbb{R}}
\newcommand*{\der}[1]{%
   \accentset{\mbox{\large\bfseries .}}{#1}}

\copyrightyear{2021} 
\acmYear{2021} 
\setcopyright{acmlicensed}\acmConference[UMAP '21]{Proceedings of the 29th
ACM Conference on User Modeling, Adaptation and Personalization}{June
21--25, 2021}{Utrecht, Netherlands}
\acmBooktitle{Proceedings of the 29th ACM Conference on User Modeling,
Adaptation and Personalization (UMAP '21), June 21--25, 2021, Utrecht,
Netherlands}
\acmPrice{15.00}
\acmDOI{10.1145/3450613.3456830}
\acmISBN{978-1-4503-8366-0/21/06}

\begin{document}

\title{Dynamic Modeling of User Preferences for Stable Recommendations}


\author{Oluwafemi Olaleke}
\email{oluwafemi.olaleke@skoltech.ru}

\author{Ivan Oseledets}
\additionalaffiliation{%
  \institution{Institute of Numerical Mathematics, Russian Academy of Sciences}
  \city{Moscow}
  \country{Russia}
}
\affiliation{%
  \institution{Skolkovo Institute of Science and Technology}
  \city{Moscow}
  \country{Russia}
}

\author{Evgeny Frolov}
\email{evgeny.frolov@skoltech.ru}
\orcid{0000-0003-3679-5311}
\affiliation{%
  \institution{Skolkovo Institute of Science and Technology}
  \city{Moscow}
  \country{Russia}
}

\renewcommand{\shortauthors}{Olaleke et al.}

\begin{abstract}
    In domains where users tend to develop long-term preferences that do not change too frequently, the stability of recommendations is an important factor of the perceived quality of a recommender system. In such cases, unstable recommendations may lead to poor personalization experience and distrust, driving users away from a recommendation service. We propose an incremental learning scheme that mitigates such problems through the dynamic modeling approach. It incorporates a generalized matrix form of a partial differential equation integrator that yields a dynamic low-rank approximation of time-dependent matrices representing user preferences. The scheme allows extending the famous PureSVD approach to time-aware settings and significantly improves its stability without sacrificing the accuracy in standard top-$n$ recommendations tasks.
\end{abstract}




\maketitle

\section{Introduction}
The typical life-cycle of a production-level recommender systems after initial training normally consists of two independent steps: generating predictions with the most recent model and retraining or incrementally updating the model as more data is collected. Ranging from minutes and hours to days and weeks, the length of intervals between subsequent (re)training steps depends on numerous factors such as frequency of updates in the product catalog, activity of users, type of algorithm, scale of the system, or computational capabilities of the infrastructure. It is often expected that the resulting model always has a consistent state and achieves the best performance. But what actually happens between the two learning steps? What is the degree of changes that the recommender system undergoes after processing newly collected data, and how does it affect the perceived quality of recommendations?

Intuitively, small changes in observations should never considerably alter the model's output (if at all), especially if these changes are aligned with the previously observed data. In contrast, more pronounced changes may significantly affect the model's predictions,
but the degree of the changes may vary between different algorithms. 
To quantify this degree of change,
one may use the notion of \emph{stability of recommendations}. As a complementary performance metric, stability is connected to such perceived qualities of a recommender system as trustworthiness, helpfulness, and informativeness \cite{Adomavicius2012Stability}. Correspondingly, the lack of consistency in recommendations, induced by a recommender system's instability, may negatively impact an overall user experience and satisfaction with the service.

Stability may not always be a desired property and certainly depends on a domain of application. For example, some online services by design may have to deal with short-lived products. It can be an online service for media recommendations with a fast expiry rate (e.g., memes) or a system of urgent same-day handyman service requests.
In such cases, a recommender system's ability to adapt to a rapidly changing environment (sometimes called plasticity \cite{Burke2007Hybrid}) may be preferred over stability. On the other side, some products may present a long-lasting value to consumers (e.g., books) and users may favor recommendations consistency over their recency or freshness.

Moreover, as shown by e.g. \citet{Hoeffler1999StablePreferences}, the dynamics of user preferences may transform depending on the environment in which products are consumed. For example, factors like difficulty of choice or experience in a product category may define whether user preferences will remain malleable or develop into a long-term form, which, in turn, may require adjusting the stability of recommendations. Hence, understanding the stability-related properties of a recommendation algorithm is a crucial task in designing a good personalization service.

In this work, we revisit the problem of generating stable recommendations and present a fresh view on the corresponding learning task. Despite significant attention to the problem, previously published works seem to almost exclusively focus on analyzing recommender systems' stability through the lens of rating prediction tasks.
Even though such an approach provides convenient instruments for modeling and analysis, a good performance in terms of rating prediction does not necessarily improve the relevance of recommendations \cite{Cremonesi2010PureSVD, Herlocker1999CF}.
As the latter plays a more important role in most real-world scenarios, we opt to focus solely on the \topn recommendations tasks.

This allows us to consider a broader family of methods that may not work well for rating prediction but achieve higher quality in terms of the relevance of recommendations. We start from a specific approach in this work, namely PureSVD, proposed by \citet{Cremonesi2010PureSVD}.
Interested in the recommendations' stability, we adopt a special dynamic learning scheme that enables a smooth transition of the model's state in time. 
Our main contributions are as follows:
\begin{itemize}
    \item we connect the problem of stable recommendations to the task of dynamic modeling of time-dependent matrices;
    \item accordingly, using the dynamical low-rank approximation theory, we extend the PureSVD approach with an incremental learning scheme aimed at generating more stable representations in time;
    \item we also present a new estimate of recommender systems stability tied to the changes in \topn recommendations lists rather than in rating prediction accuracy.
\end{itemize}
Our experiments demonstrate that the proposed approach significantly improves the stability of recommendations comparing to its non-dynamic predecessor. We also observe that higher levels of stability can be achieved without deteriorating the relevance of recommendations. The entire source code for reproducing our work is available online\footnote{https://github.com/evfro/Stable\_Recommendations}.
\section{Related work}\label{sec:related}
According to \citet{Adomavicius2010Stability, Adomavicius2012Stability}, stability of recommendations characterizes the degree to which the recommender system's predictions remain consistent as more training data arrives.
In turn, predictions are considered inconsistent if they change even when newly added data is in complete agreement with the history of past observations. The latter condition was explicitly assumed by the authors in their analysis, meaning that \emph{new observations must at least approximately follow the same distributional law as the past data}. We note that the work focused specifically on the rating prediction task. Consequently, the measure of inconsistencies was defined as the difference between predicted ratings on some target items in two subsequent learning steps.

Continuing this line of work, \citet{Adomavicius2015MetaStability} developed a stability-maximizing technique based on a meta-algorithmic approach. They proposed to use bootstrap aggregation (bagging) with uniform voting followed by an iterative smoothing procedure explicitly designed to compensate for remaining instabilities. Acknowledging the computational bottleneck of such an approach, the authors proposed an approximate scheme that restricted the number of models built at each iteration. The resulting solution was demonstrated to improve the stability of recommendations along with the rating prediction accuracy.

\citet{Shriver2019StabilityEval} proposed a slightly different view on recommendations stability. The authors defined it simply as the measure of consistency of recommendations, given that changes to the training data are small. Hence, they \emph{relax the data agreement assumption} introduced in \cite{Adomavicius2010Stability}.
Accordingly, if a small number of changes leads to a disproportionately large shift in recommendations, the system is considered unstable.
The authors also noted that precisely determining the stability is problematic even for small-sized problems, as the search space for exploring all possible variations in input data quickly becomes infeasible. Hence, they proposed several heuristic models of influence that helped narrow down the search space and identify major instability factors for different types of algorithms

In contrast to \cite{Adomavicius2010Stability, Adomavicius2012Stability}, the degree of instability in \cite{Shriver2019StabilityEval} was estimated through the comparison of lists of recommendations rather than rating predictions. 
For that purpose, the authors introduced a new \emph{TopOut} metric. It is based on the assumption that top-ranked items in recommendation lists are often the most difficult to change. If, after retraining, such items disappear from recommendations, it signals about a considerable shift in the model state and indicates instability.
The main goal of the work was to explore the notion of stability via possible mechanisms of inducing significant changes to algorithms' predictions. Hence, the authors only evaluated instability and did not provide any information about the actual performance with respect to conventional evaluation metrics. 



\section{Proposed approach}
In this section, we tie our definition of the recommendations stability to specific optimization problem formulation. Using the dynamical low-rank approximation theory for time-dependent matrices \cite{Lubich2014ProjectorSplitting}, we combine this formulation with the PureSVD approach \cite{Cremonesi2010PureSVD}, which in turn enables the dynamic modeling of user preferences and provides a smooth transition between the subsequent learning steps. Our main goal is to develop an efficient approach that makes recommendations more stable without sacrificing their relevance.

\subsection{Estimating recommendations stability}\label{sec:stability}
Following the definition of stability proposed by \citet{Shriver2019StabilityEval}, \emph{we relax the data agreement assumption} as well.
That way, it seems to be more realistic as there is generally no control over the nature of incoming data. Moreover, identifying data alignment is an extremely challenging task when observations are missing not at random, as the governing mechanisms for data generation are not fully understood.
Continuing the provided in \cite{Shriver2019StabilityEval} reasoning, we also only require that \emph{the amount of new observations used to update the model is relatively small}. This assumption helps us formulate a proper optimization objective described further in \Sref{sec:method}.

We additionally note that focusing on rating prediction to evaluate the stability of recommendations can be misleading. Assuming that users are generally less interested in the low-rated items than in the top-rated ones, recommendations inconsistency over the latter will likely have a more pronounced negative impact on user expectations. 
Hence, a similar line of arguments used against the error-based metrics \cite{Cremonesi2010PureSVD,Herlocker1999CF} can be applied here as well.
Indeed, suppose an algorithm remains consistent only in the lower spectrum of ratings but not in the higher ratings part. In that case, the calculated stability metric can still be very high if the fraction of the top-rated items is small (which is a typical case in practice). This would lead to a paradoxical situation as the calculated metric would be the opposite of the actual user experience.



To avoid such pitfalls, \emph{we estimate the consistency of predictions based on the similarity between two ordered lists of \topn recommendations generated in subsequent time-steps}. Higher similarity values will indicate higher consistency of recommendations and, therefore, better stability. As mentioned in \Sref{sec:related}, removing the top-ranked item from the list is likely to have a more pronounced effect on the perceived recommendations consistency. We generalize this idea and \emph{penalize any reordering of items based on their position in the list}. For example, moving a top-ranked item to the end of the list will be penalized more than moving the second-to-last item there.

For this task, we calculate similarities based on the \emph{Weighted Jaccard Index} (WJI) \cite{Chierichetti2010WeightedJaccard}, using the analogy with the task of finding weighted similarity between the text documents, as described in e.g. \cite{Aryal2019WeightedJaccard}. We encode recommendation lists as ``bag-of-items'' (BOI) vectors $\bvec{x} = \left( x_i \right)_{i=1}^N$, where $x_i$ denotes the rank (position in the list) of item $i$ and $N$ is the size of entire item catalog. The \topn similarity between two recommendations lists is measured as the WJI of their BOI representations $\bvec{u}, \bvec{v}$:
\begin{equation}\label{eq:wji}
    \simi(\bvec{u}, \bvec{v})\,@\,n =\frac{\sum_{i=1}^{N} \min \left\{w_{i}(\bvec{u}), w_{i}(\bvec{v})\right\}}{\sum_{i=1}^{N} \max \left\{w_{i}(\bvec{u}), w_{i}(\bvec{v})\right\}},
\end{equation}
where weights $w_{i}(\bvec{x}) = x_i^{-1}$ are calculated as the reciprocal rank of item $i$, and $w_{i}(\bvec{x})=0$ if $x_i > n$. Note that the values of this similarity are non-negative: \mbox{$0 \leq \simi(\bvec{u}, \bvec{v})\,@\,n \leq 1$}.


\subsection{Dynamic modelling of user preferences}\label{sec:method}

In \Sref{sec:stability}, we established the definition of stability using the assumption of small changes between consequent learning steps. Accordingly, we say that the system lacks stability if small changes in data induce disproportionately large changes in the recommender model. In order to be stable, the model must adequately respond to the updates in training data.
In the case of PureSVD approach, adding more data requires recomputing the entire model. As the solution is based on standard SVD, it only guarantees the best possible low-rank approximation of the data. However, it imposes no "proximity" constraint between the updated model and its previous state. As the newly added data may, in general, have an arbitrary nature, the degree of changes in the model can potentially be large, leading to significant inconsistencies. We aim to mitigate this problem with the help of a special formulation of the optimization objective.

Let $\bvec{A}(t) \in \R^{M \times N}$ denote a matrix of interactions between $M$ users and $N$ items at time step $t$, and $\bvec{Y}(t) \in \R^{M \times N}$ denote its low-rank approximation at the same time step. For simplicity, we assume that the number of rows and columns stay fixed across all time steps.
According to the PureSVD approach, all the values for missing interactions in $\bvec{A}(t)$ are set to 0. At any fixed moment of time the following minimization problem is solved:
\begin{equation}
    \min_{\bvec{Y}(t):\, \text{rank}(\bvec{Y}(t))=r} \|\bvec{A}(t)-\bvec{Y}(t)\|,
\end{equation}
where $\| \cdot \|$ denotes the Frobenius norm. The solution is given as
\begin{equation}
    \bvec{Y}(t) = \bvec{U}(t)\;\bvec{\Sigma}(t)\;\bvec{V}(t)^\top,
\end{equation}
where latent feature matrices $\bvec{U}(t) \in \R^{M \times r}$ and $\bvec{V}(t) \in \R^{N \times r}$ each have $r$ orthonormal columns and $\Sigma(t) \in \R^{r \times r}$ is a diagonal matrix of singular values. Clearly, there is \emph{no restriction on the rate of change} at different time steps in these matrices. \emph{We aim to fix this issue by the virtue of dynamical modeling approach}, described next.

\noindent\begin{minipage}[t]{.60\textwidth}%
\begin{algorithm}[H]
  \SetKwInOut{Input}{Input}
  \SetKwInOut{Output}{Output}
  \Input{\quad $\Delta\bvec{A}$, \Comment{new data collected since the last time-step}\\
  \quad $\bvec{U}_0, \bvec{S}_0, \bvec{V}_0$ \Comment{last state of the model at time $t-1$}}
  \Output{\quad $\bvec{U}_1, \bvec{S}_1, \bvec{V}_1$ \Comment{new state of the model at time $t$}}
    $\bvec{K}_{1} \leftarrow \bvec{U}_{0}\bvec{S}_{0} + \Delta\bvec{A} \bvec{V}_{0}$ \\
    $\bvec{U}_{1},\,\widehat{\bvec{S}}_{1} \leftarrow \QR{\bvec{K}_{1}}$ \Comment{QR or SVD decomposition of $\bvec{K}_{1}$}\\
    $\widetilde{\bvec{S}}_{0} \leftarrow \widehat{\bvec{S}}_{1} - \bvec{U}_{1}^\top \Delta\bvec{A} \bvec{V}_{0}$ \\
    $\bvec{L}_{1} \leftarrow \bvec{V}_{0}\widetilde{\bvec{S}}_{0}^\top + \Delta \bvec{A}^\top \bvec{U}_{1}$ \\
    $\bvec{V}_{1},\,\bvec{S}_{1}^\top \leftarrow \QR{\bvec{L}_{1}}$ \Comment{QR or SVD decomposition of $\bvec{L}_{1}$}\\
  \caption{One step of the PSI algorithm}
  \label{alg:projector-spitting}
\end{algorithm}
\end{minipage}%
\begin{minipage}[t]{.40\textwidth}%
\vspace*{\baselineskip}
\begin{table}[H]
    \caption{Datasets statistics after pre-processing.}
    \centering{
    \begin{tabular}{l*{4}{c}}
    \hline
      & \textbf{\# users} & \textbf{\# items} & \textbf{density}\\
    \hline
    \textbf{ML-1M} & 6038 & 3533 & 2.70\% \\
    \textbf{AMZvg} & 8966 & 16370 & 0.11\% \\
    \textbf{AMZb} & 27746 & 8643 & 0.03\% \\
    \hline
    \end{tabular}
    }
    \label{tab:datasets}
\end{table}
\end{minipage}%

\medskip

More specifically, according to our definition of stability, \emph{the degree of change in $\bvec{Y}(t)$ at each step must be proportionate to that of $\bvec{A}(t)$}. One can formally write this requirement in terms of a \emph{dynamical low-rank approximation} objective \cite{Koch2007DLRA}:
\begin{equation}\label{eq:timeopt}
    \min_{\bvec{Y}(t):\, \text{rank}(\bvec{Y}(t))=r} \|\der{\bvec{A}}(t)-\der{\bvec{Y}}(t)\|,
\end{equation}
where $\der{\bvec{A}}=\frac{d}{dt}\bvec{A}$.
The objective in \eqref{eq:timeopt} ensures that $\bvec{Y}(t)$ changes smoothly with small changes in $\bvec{A}(t)$. We expect that it will improve consistency between the recommendations generated in subsequent time-steps.
The objective \eqref{eq:timeopt} yields a differential equation on the manifold of matrices of a given rank $r$, which can be solved numerically in a fully explicit way using the Projector-Splitting Integrator (PSI) method proposed by \citet{Lubich2014ProjectorSplitting}. PSI offers an \emph{incremental scheme} that takes only the newly added data to directly update the model, which \emph{eliminates the need to recompute the entire model from scratch} inevitable in the case of the standard PureSVD approach. See \Aref{alg:projector-spitting} for implementation details. At $t=0$ step the model $\bvec{Y}(0)$ is obtained with PureSVD. We note that \emph{the resulting solution inherits all the practical advantages and benefits of the PureSVD approach}\footnote{A brief overview of some of the advantages can be found at https://www.eigentheories.com/blog/to-svd-or-not-to-svd/}.

\section{Experiments}
In this section, we present an experimental setup aimed at evaluating the performance of the PSI-based approach against its non-incremental PureSVD-based counterpart in the standard \topn recommendations task. PureSVD is an SVD-based model that conveniently represents users as a combination of item features by means of an orthogonal projection in the latent feature space. It presents a strong baseline and outperforms several more sophisticated models in \topn recommendation tasks \cite{Cremonesi2010PureSVD}. We adapt PureSVD to our incremental learning setup by iteratively recomputing it on the entire previous history up to the current step.

\subsection{Datasets}
We evaluate our method on several publicly available datasets: \emph{MovieLens-1M} (ML1M) \cite{harper2015movielens}, \emph{Amazon Beauty} (AMZb), \emph{Amazon Video Games} (AMZvg) \cite{mcauley2015image}. In all datasets,  we select only interactions with the rating of at least 4, and since we are not interested in predicting ratings, we binarize them. For the Amazon datasets, we set a threshold for the selection of users who have interacted with a particular number of items. Specifically, for AMZvg  we filtered out all users who have interacted with less than 10 items, while for AMZb, we filtered out users who have interacted with less than 2 items. No filtering was performed on items, i.e., we do not require any specific number of user ratings per item. The main characteristics of the datasets after preprocessing are provided in Table \ref{tab:datasets}.

\subsection{Evaluation Methodology}
We implement a \emph{stepwise data splitting} method for all datasets based on time increments.
For the initial training step, we take all but the last 8 months of data. This ensures that all models have a reliable initial state. The remaining 8 months are split into time intervals of equal size. The number of distinct time steps is determined based on user activity so that the overall number of interactions in each step is not too high or too low. We found that one-month intervals were sufficient, resulting in exactly 8 time steps. Subsequently, at each step, we hide the most recently consumed item for every user who has more than one interaction within the current month-long interval. The hidden items form a holdout dataset and are used to evaluate the quality of recommendations. The remaining interactions are either merged with the training set (for PureSVD) or form an incremental update matrix $\Delta \bvec{A}$ (serving as input for PSI in Algorithm \ref{alg:projector-spitting}).

At each step, we perform the standard evaluation of \topn recommendations. We generate a list of top-ranked items for every test user based on their currently known preferences and evaluate it against the holdout. We iteratively measure and record the quality of recommendations at each time step using several standard evaluation metrics: \emph{HitRate} (HR), \emph{Mean Reciprocal Rank} (MRR), and \emph{Coverage}. The latter is computed simply as the fraction of all unique items recommended by an algorithm against the total amount of unique items from the training set.

We also evaluate the stability of recommendations using the WJI score, as defined in \Eref{eq:wji}. Note that unlike other metrics that are computed within a single step, stability is measured by comparing \topn recommendations lists generated in two subsequent time-steps. For all metrics, the results are averaged across all test users. The final scores are obtained by averaging across all time intervals.

\medskip





\begin{figure*}[ht]
    \centering
    \subfigure[][]{%
        \label{fig:mlres}
        \includegraphics[width=\textwidth]{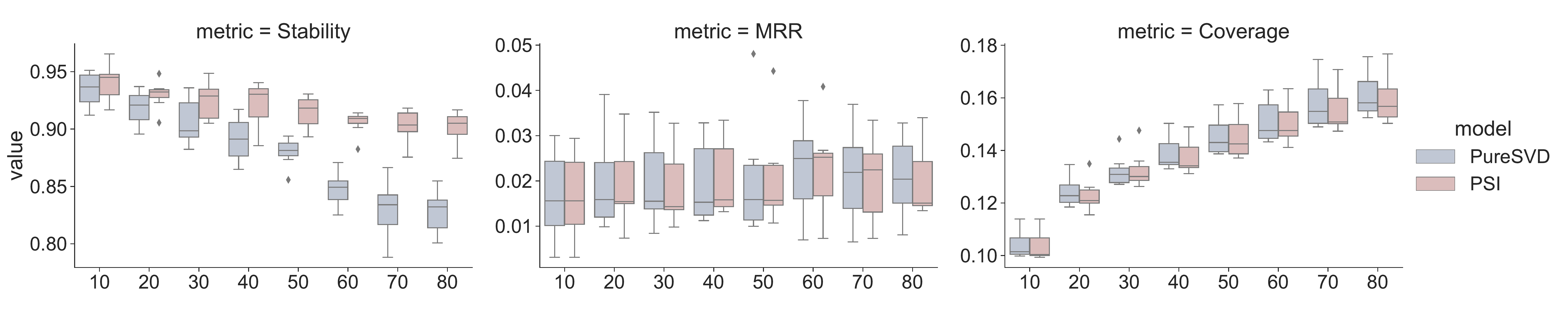}}
    \hfill%
    \subfigure[][]{%
        \label{fig:amzgres}
        \includegraphics[width=\textwidth]{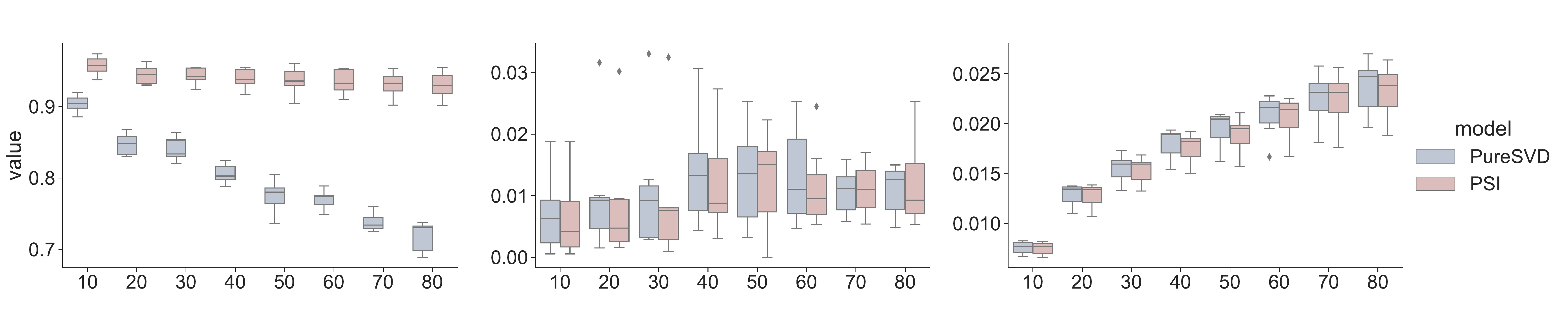}}
    \subfigure[][]{%
        \label{fig:amzbres}
        \includegraphics[width=\textwidth]{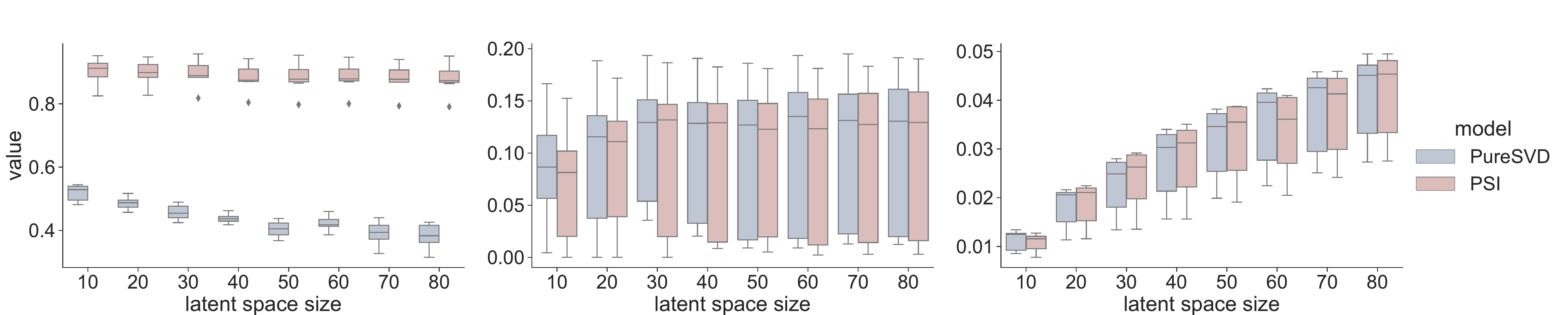}}
    \caption{Experimentation results for \subref{fig:mlres} Movielens-1M, \subref{fig:amzgres} Amazon Video Games, and \subref{fig:amzbres} Amazon Beauty datasets evaluated against three metrics going from left to right: Stability, MRR, Coverage. We do not report HR metric results as they are similar to MRR.}
    \label{fig:results}
    \Description[Comparing PureSVD and PSI models.]{<All experiments.>}
\end{figure*}
\section{Results}
The results of our experiments are demonstrated in \fref{fig:results}. We do not report HR scores as they highly correlate with MRR. However, the complete set of measurements can be found in our online repository. Overall, PSI demonstrates higher stability in all experiments, while PureSVD becomes significantly more inconsistent as the dimensionality of the latent space grows. Note that higher numbers of latent features are likely to provide better personalization as more subtle variations in user behavior are taken into account. Conversely, a smaller latent space induces a strong popularity bias, which may explain why the differences in stability between the two models tend to decrease in that case.

Interestingly, the striking difference in stability between PureSVD and PSI does not result in significant changes in MRR, which may seem unreasonable at first glance as both metrics depend on the ordering of items. We hypothesize that this is due to an irreversible nature of stability. For example, imagine two test users whose holdout items were correctly predicted at the first and second positions of their \topn recommendations lists respectively. If we flip the positions in both lists so that the first item becomes the second and vice versa, the average MRR score will remain unchanged. The degraded ranking for one user gets compensated by the improved ranking for another one. In turn, the stability can only decrease for both users as the recommendations lists are not the same after the flip.

One may expect that the decreased stability of PureSVD could at least result in a much higher catalog coverage. Indeed, higher inconsistencies significantly change the ordering within \topn recommendations lists, which potentially may promote some new items that were unseen before. However, as we can see from the third column of \fref{fig:results}, in most cases, there is no statistically significant difference in coverage between PureSVD and PSI, and the overall effect is negligible. Worth noting here that the quality of SVD-based techniques, including coverage, can be significantly boosted with a simple data-normalization technique \cite{Nikolakopoulos2019Eigenrec,Frolov2019HybridSVD}. Hence, when both algorithms are properly configured, the differences in coverage may potentially become unnoticeable. Adopting this technique to dynamic settings of PSI is an interesting problem that we leave for further research.

Overall, the observed results are in agreement with our initial assumption that optimizing the objective \eqref{eq:timeopt} makes the transition of the model between subsequent time-steps more consistent, which yields better stability.



\section{Conclusions}
We provide a fresh view on the definition of recommendations stability. We show how to formulate in terms of a dynamical low-rank approximation of a time-dependent user preferences matrix. It enables an incremental learning scheme based on the PureSVD approach and ensures a smooth transition of the model between subsequent time-steps. To assess the effect of this scheme on the stability properties of a recommender system, we also propose a new measure of recommendations consistency that avoids unrealistic assumptions on the nature of observations data. We connect it to the \topn recommendations task instead of commonly used rating prediction, which makes it more practical in many real-world scenarios. The calculation is based on finding the weighted similarity between ordered lists of recommended items. Our experimental results indicate that the proposed approach achieves higher stability over its non-incremental counterpart without sacrificing the general performance in terms of standard relevance- and ranking-based metrics.

This work is an initial step towards more general dynamic modeling solutions, and the proposed approach can be improved in several ways. For example, an additional dynamic normalization of data may lead to better coverage and overall performance. Another interesting direction is to combine this approach with techniques that allow adding new users and items on-the-fly with the minimal distortion to the approximation quality. It can be naturally extended to general context-aware settings using tensor factorization. We leave these topics for our future research.


\bibliographystyle{ACM-Reference-Format}
\bibliography{main}

\end{document}